\documentclass[conference]{IEEEtran}
\IEEEoverridecommandlockouts

\usepackage{cite}
\usepackage{amsmath,amssymb}
\usepackage{graphicx}
\usepackage{booktabs}
\usepackage{multirow}
\usepackage{array}
\usepackage{xcolor}
\usepackage{placeins}
\usepackage{stfloats}
\usepackage{balance}
\usepackage{url}
\usepackage{hyperref}
\hypersetup{hidelinks}
\begin{document}

\title{Vision-Guided Text Prompt Tuning for Multimodal Sentiment Analysis}
\author{
\IEEEauthorblockN{
Xiaoran Kou\textsuperscript{1},
Jingyi Wu\textsuperscript{2},
Peng Sun\textsuperscript{3}\textsuperscript{*},
Yang Liu\textsuperscript{1}\textsuperscript{*},
Hong Chen\textsuperscript{1}
}
\IEEEauthorblockA{
\textsuperscript{1}College of Electronic and Information Engineering, Tongji University, Shanghai 201804, China\\
\textsuperscript{2}College of Intelligent Robotics and Advanced Manufacturing, Fudan University, Shanghai 200433, China\\
\textsuperscript{3}Division of Natural and Applied Sciences, Duke Kunshan University, Suzhou 215316, China\\
yangliu25@tongji.edu.cn,\quad
chenhong2019@tongji.edu.cn,\quad
jingyiwu23@m.fudan.edu.cn,\quad
peng.sun568@duke.edu
}
\thanks{\textsuperscript{*}Corresponding authors

This work was supported by DKU Professional Discretionary Funds 26AKUG0088 and 00AKUG0343.}
}
\maketitle

\begin{abstract}
Multimodal sentiment analysis requires effective modeling of both verbal semantics and non-verbal affective cues. 
A central challenge is to calibrate text-centered sentiment understanding with visual facial evidence in a controlled, adaptive, and parameter-efficient manner. 
Text usually serves as the semantic anchor, whereas visual cues provide complementary evidence for ambiguous or implicit expressions; however, indiscriminate fusion may introduce visual noise and distort textual semantics. 
Moreover, fully fine-tuning large visual and textual encoders is costly and prone to overfitting on limited and scenario-dependent MSA benchmarks. 
To address these issues, we propose Vision-Guided Text Prompt Tuning (VG-TPT), which formulates visual-text sentiment modeling as controllable visual calibration of frozen text representations. 
VG-TPT injects visual affective cues into a frozen BERT encoder through layer-wise adaptive prompts, rather than relying on late-stage feature fusion or full backbone tuning. 
A co-guided router composes prompts from a trainable prompt bank according to both the evolving text state and the visual guidance feature, enabling sample-specific and layer-specific modulation. 
Experiments on CMU-MOSEI and CMU-MOSI show that VG-TPT consistently improves over text-only baselines and achieves competitive or superior performance compared with several full-modality methods, while updating only 2.4M trainable parameters.
The code is available at \href{https://github.com/ma-tubu/VG-TPT}{\texttt{https://github.com/ma-tubu/VG-TPT}}.
\end{abstract}


\section{Introduction}

Multimodal sentiment analysis (MSA) aims to infer human affective states from heterogeneous multimedia signals, including language, vision, and acoustics~\cite{zadeh2017tfn,tsai2019mult}. 
Beyond benchmark evaluation, MSA is increasingly required in practical multimedia systems such as embodied perception, autonomous driving, human--robot interaction, and affect-aware human--computer interfaces~\cite{zuo2026embodied,yang2023aide,liu2024gvaed,liu2024ampnet}. 
Such application-oriented MSA requires efficient adaptation across users, environments, devices, and domain-specific distributions, making full fine-tuning of large modality encoders costly and potentially unstable under limited paired annotations. 
Prompt tuning provides a parameter-efficient adaptation interface that preserves pretrained unimodal knowledge while enabling task- and modality-specific guidance. 
In this context, text usually serves as the semantic anchor, whereas visual signals provide non-verbal affective evidence such as facial expressions and gestures.

\begin{figure}[!t]
    \centering
    \includegraphics[width=\linewidth]{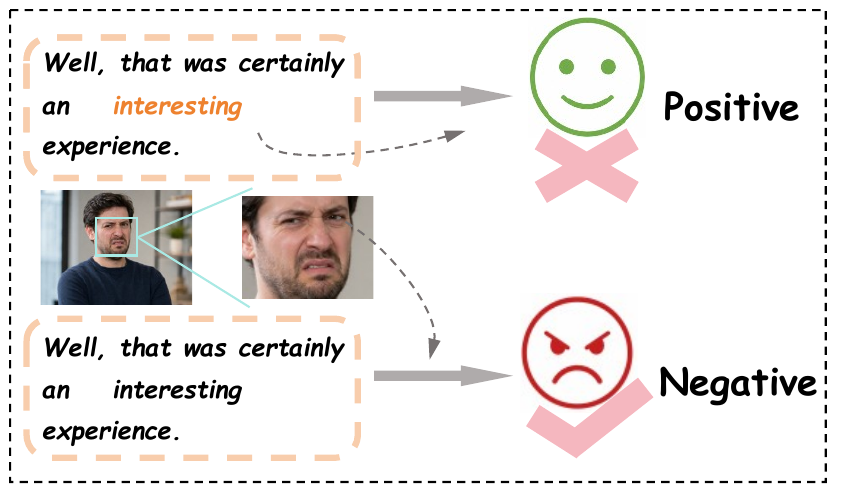}
    \caption{
    Motivation of VG-TPT.
    Text alone may lead to an ambiguous or incorrect sentiment prediction, while facial expressions provide fine-grained affective cues to refine and disambiguate text-centered sentiment interpretation.
    }
    \label{fig:motivation}
\end{figure}

Although text is often the dominant modality in sentiment analysis, its dominance also brings a structural limitation~\cite{lei2024tmrn}. 
Text is usually semantically dense and easier to model, while visual cues are often sparse, noisy, and highly dependent on the speaker and context~\cite{lei2024tmrn,he2026ebmc}. 
As a result, existing multimodal models may over-rely on textual polarity and under-utilize non-verbal affective evidence~\cite{he2026ebmc}. 
However, in many real-world utterances, the decisive sentiment is not explicitly encoded in words but is revealed by facial expressions. 
As illustrated in Fig.~\ref{fig:motivation}, an utterance such as ``Well, that was certainly an interesting experience'' may appear neutral or mildly subjective from text alone, while a disgusted facial expression clearly indicates negative sentiment. 
The preceding example suggests that text should serve as the semantic anchor, but visual facial cues are necessary to calibrate, refine, and sometimes overturn text-centered sentiment interpretation~\cite{wang2024mgr3net,wang2024fer,yang2022context}.

Conventional multimodal fusion further exposes a limitation in visual-text sentiment modeling. 
Most methods introduce visual information through feature concatenation, cross-modal attention, or decision-level fusion after independent modality encoding~\cite{zadeh2017tfn,zadeh2018mfn,tsai2019mult,han2021mmim,yang2022contextual}. 
Although these paradigms model cross-modal correlations, vision is usually treated as a parallel auxiliary stream and fused with text only at a restricted stage. 
Late-stage fusion can adjust the final representation or prediction, but has limited ability to participate in the formation of textual semantics. 
A more principled strategy should allow visual evidence to intervene in the internal evolution of text representations across Transformer layers.

Visual guidance also exhibits sample-dependent reliability~\cite{he2026ebmc}. 
Some samples require mild visual calibration, whereas others need strong correction when text and facial expressions conflict. 
As textual states change across layers, visual guidance should also vary with the current semantic level. 
However, conventional prompt-based fusion methods typically employ globally shared prompts, making them inadequate for sample-specific and layer-specific variations~\cite{lester2021prompt,liang2022promptfusion,jia2022vpt}. 
Motivated by these observations, VG-TPT dynamically composes prompts from a trainable prompt bank conditioned on both the evolving text state and visual guidance feature.

To address these challenges, we propose VG-TPT for multimodal sentiment analysis. 
Built upon a frozen visual encoder and a BERT-based textual backbone, VG-TPT introduces three key designs: 
(1) a text-anchored visual guidance scheme that injects facial affective cues into the internal layers of the text encoder instead of performing late feature concatenation; 
(2) a layer-wise co-guided prompt routing mechanism that generates prompts according to both the evolving text state and the visual guidance feature, enabling adaptive modulation across different semantic layers; 
and (3) a routing-guided prompt composition strategy that selectively activates and combines prompt bases from a trainable prompt bank using learned routing weights, improving the adaptability of prompt-based visual-text fusion.

Our primary contributions are summarized as follows:
\begin{itemize}
    \item We propose VG-TPT to formulate visual-text sentiment modeling as controllable visual calibration of text-centered representations, enabling facial cues to correct ambiguous textual sentiment cues.

    \item We introduce a layer-wise co-guided prompt routing mechanism that dynamically composes prompts from a trainable prompt bank, providing sample-specific and layer-specific visual modulation.

    \item We achieve parameter-efficient visual-text adaptation by freezing the visual encoder and BERT backbone. With only 2.4M trainable parameters, VG-TPT outperforms text-only baselines and remains competitive with several full-modality methods.
\end{itemize}

\section{Related Work}

\subsection{Multimodal Sentiment Analysis}

Multimodal sentiment analysis (MSA) integrates language, visual, and acoustic cues to infer human sentiment~\cite{zadeh2017tfn,tsai2019mult}. Existing methods have evolved from simple feature-level fusion, such as concatenation and tensor fusion~\cite{zadeh2017tfn,zadeh2018mfn}, to attention-based fusion, multimodal Transformers~\cite{tsai2019mult}, representation disentanglement, and dynamic modality weighting~\cite{han2021mmim,lei2024tmrn,wang2025dlf,zhang2025msfn,li2024corrkd,jiang2025ddse,wu2025deva,xu2025hcm}. Although these approaches effectively model cross-modal complementarity, they usually treat modalities as parallel streams and fuse them after independent encoding. As a result, visual information is often used as an auxiliary feature rather than a fine-grained guidance signal for text understanding. Unlike these methods, we focus on vision-guided text modeling, where facial cues are injected into the textual encoder through layer-wise adaptive prompts to refine and disambiguate text-centered sentiment representations.

\FloatBarrier

\begin{figure*}[!t]
    \centering
    \includegraphics[width=0.90\textwidth]{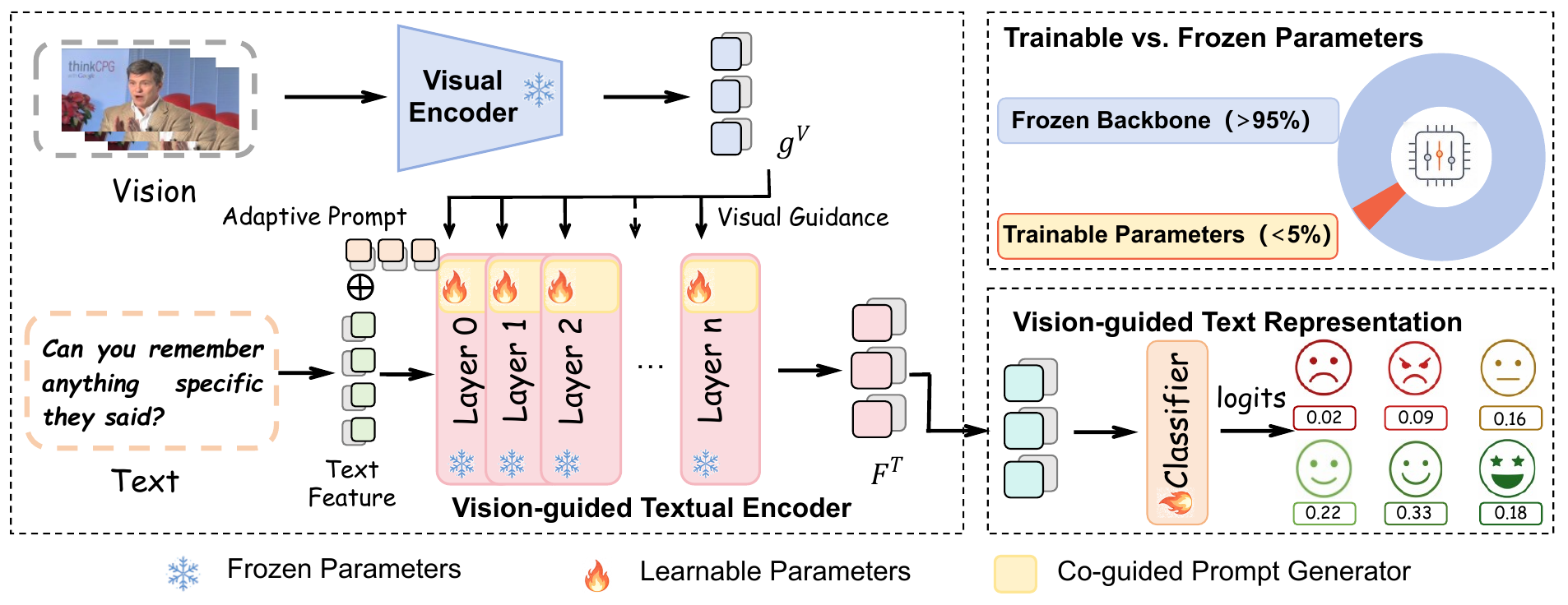}
    \caption{
    Overall architecture of VG-TPT.
    A frozen visual encoder extracts the visual guidance feature $g^V$, which is injected into the BERT-based textual encoder through layer-wise adaptive prompts.
    The final vision-guided text representation is used for sentiment prediction.
    }
    \label{fig:overall_framework}
\end{figure*}

\subsection{Prompt Tuning for Multimodal Fusion}

Prompt tuning freezes pretrained backbones and optimizes only a small number of continuous prompt tokens, providing a parameter-efficient alternative to full fine-tuning~\cite{lester2021prompt,jia2022vpt}. Recent prompt-based multimodal fusion methods use learnable prompts to connect separately pretrained unimodal encoders and inject cross-modal information into a designated main modality encoder~\cite{liang2022promptfusion}. However, conventional prompt fusion often adopts globally shared prompts, which are applied to all samples regardless of their specific multimodal context. Such globally shared prompting limits prompt adaptivity and expressiveness, especially for sentiment analysis where different text-visual pairs may require different affective interpretation patterns. Inspired by recent adaptive prompt fusion studies, we introduce a layer-wise co-guided prompt generation mechanism. Our router jointly considers the current text state and visual guidance feature at each prompted Transformer layer, and selectively composes adaptive prompts from a trainable prompt bank, enabling sample-specific and layer-specific visual modulation with few trainable parameters.

\section{Method}
\label{sec:method}

\subsection{Overview}
\label{subsec:overview}

Given a text utterance $X^T$ and its corresponding visual input $X^V$, our goal is to predict the sentiment intensity or sentiment category $y$. 
Unlike conventional multimodal fusion methods that combine textual and visual representations only at the feature or decision level, VG-TPT adopts a text-anchored vision-guided framework. 
In our formulation, the text modality serves as the primary semantic anchor, while the visual modality provides fine-grained affective cues, such as facial expressions, to refine and disambiguate text-centered sentiment understanding.

As shown in Fig.~\ref{fig:overall_framework}, VG-TPT first uses a frozen visual encoder to extract a compact visual guidance feature:
\begin{equation}
    g^V = E_V(X^V),
\end{equation}
where $E_V(\cdot)$ denotes the visual encoder. 
The text input $X^T$ is encoded by a BERT-based textual encoder. 
Instead of directly concatenating $g^V$ with the final text feature, we inject visual guidance into the textual encoder through layer-wise adaptive prompts. 
The resulting vision-guided text representation $F^T$ is then fed into a prediction head:
\begin{equation}
    \hat{y} = C(F^T),
\end{equation}
where $C(\cdot)$ denotes the sentiment prediction head.

\subsection{Layer-wise Vision-guided Text Encoding}
\label{subsec:layerwise_encoding}

Let $H_l^T \in \mathbb{R}^{B \times N \times d}$ denote the text hidden states before the $l$-th Transformer layer, where $B$ is the batch size, $N$ is the text sequence length, and $d$ is the hidden dimension. 
For each prompted layer $l \in \mathcal{S}$, where $\mathcal{S}$ denotes the set of layers equipped with vision-guided prompts, we generate a prompt sequence $P_l$ conditioned on both the current textual state and the visual guidance feature.

The generated prompt is first inserted after the \texttt{[CLS]} token:
\begin{equation}
    \widetilde{H}_l^T =
    [h_{\mathrm{cls}}^l; P_l; h_1^l; \cdots; h_N^l],
\end{equation}
where $h_{\mathrm{cls}}^l$ denotes the \texttt{[CLS]} representation at layer $l$.

The augmented sequence is then processed by the Transformer layer, after which the inserted prompt tokens are removed:
\begin{equation}
    H_{l+1}^T =
    \mathrm{RemovePrompt}
    \left(
    \mathrm{Transformer}_l
    \left(
    \widetilde{H}_l^T
    \right)
    \right).
\end{equation}

The transient prompt injection strategy allows visual affective cues to interact with textual tokens through self-attention, while preserving the original text sequence structure across layers. 
Compared with one-shot input-level or output-level fusion, the proposed layer-wise design enables visual guidance to progressively modulate textual representations at different semantic levels.

\subsection{Co-guided Prompt Routing}
\label{subsec:routing}

A fixed prompt treats all samples equally, which is insufficient for multimodal sentiment analysis where different text-visual pairs may require different affective interpretation patterns. 
To address this issue, we introduce a co-guided prompt routing mechanism, where the prompt activation is jointly determined by the current textual state and the visual guidance feature.

For the $l$-th prompted layer, the current textual state and visual guidance feature are projected into two routing subspaces. 
Specifically, $W_T$ and $W_V$ project them into $d_T$- and $d_V$-dimensional routing subspaces, respectively. 
The joint routing condition is formulated as:
\begin{equation}
    z_l =
    [u_l^T; u^V]
    =
    [W_T h_{\mathrm{cls}}^l; W_V g^V].
\end{equation}
Here, $u_l^T \in \mathbb{R}^{B \times d_T}$ and $u^V \in \mathbb{R}^{B \times d_V}$ denote the text-side and vision-side routing embeddings, respectively. 
Thus, $z_l \in \mathbb{R}^{B \times d_r}$ with $d_T+d_V=d_r$.

We compare $z_l$ with a set of frozen routing anchors to obtain the routing distribution:
\begin{equation}
    \alpha_l =
    \mathrm{Softmax}\left(\frac{z_l A}{\tau}\right).
\end{equation}
Here, $A \in \mathbb{R}^{d_r \times K}$ denotes the routing anchors, $K$ is the number of prompt bases, and $\tau$ is a temperature coefficient. 
The resulting routing distribution $\alpha_l \in \mathbb{R}^{B \times K}$ determines how strongly each prompt basis should be activated.

\begin{figure*}[!t]
    \centering
    \includegraphics[width=0.90\textwidth]{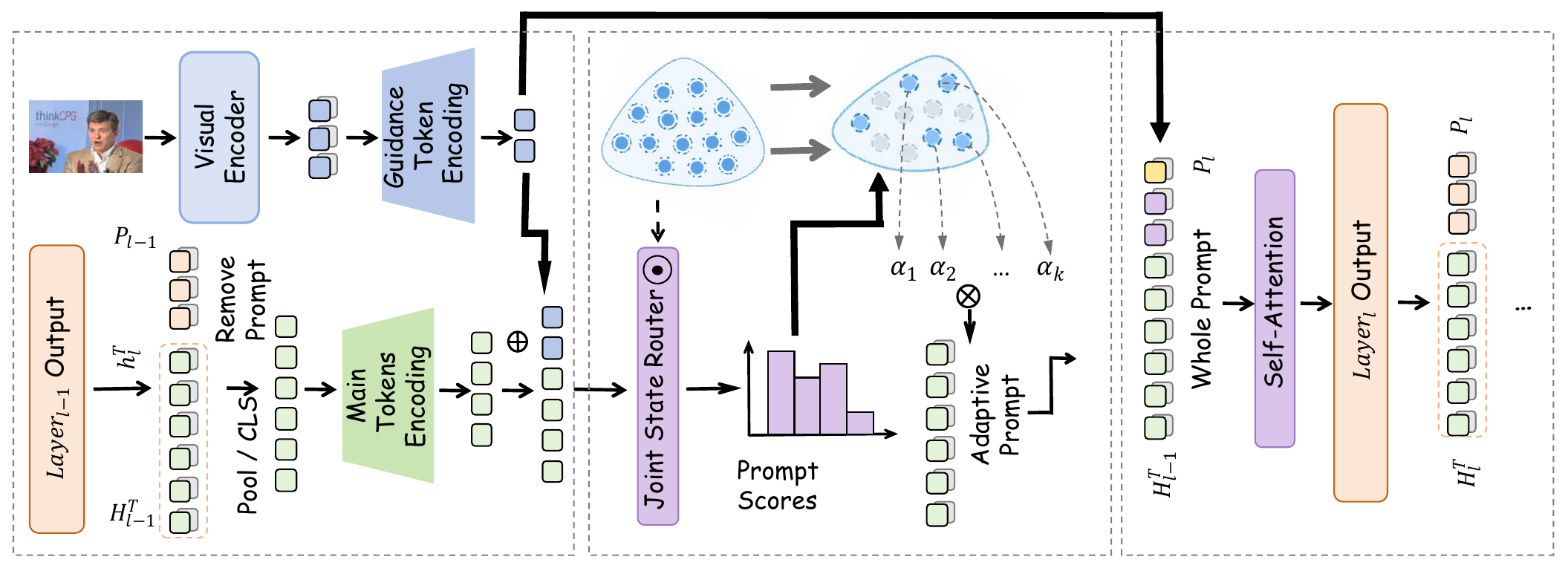}
    \caption{
    Illustration of the co-guided adaptive prompt generation mechanism.
    The current text state and the visual guidance feature are jointly used to produce routing scores over a trainable prompt basis bank.
    The routing scores softly activate different prompt bases and compose a layer-specific adaptive prompt.
    }
    \label{fig:prompt_generator}
\end{figure*}

As illustrated in Fig.~\ref{fig:prompt_generator}, this mechanism makes prompt selection both text-aware and vision-aware. 
The current text state provides information about what the textual encoder has already captured, while the visual guidance feature provides complementary affective evidence. 
Therefore, the model can adaptively select suitable prompts according to the text-visual context of each sample and each layer.

\subsection{Routing-guided Adaptive Prompt Composition}
\label{subsec:prompt_composition}

Based on the routing distribution, we compose an adaptive prompt from a trainable prompt basis bank. 
For each prompted layer, we maintain a set of learnable prompt bases:
\begin{equation}
    \mathcal{B}_l =
    \{B_{l,k}\}_{k=1}^{K},
    \quad
    B_{l,k} \in \mathbb{R}^{M \times d},
\end{equation}
where $B_{l,k}$ denotes the $k$-th prompt basis at layer $l$, and $M$ is the length of the adaptive prompt.

Instead of selecting a single prompt basis, we perform soft prompt activation and weighted prompt composition. 
To explicitly preserve the visual guidance signal, we further project the visual feature into a visual instruction token. 
The final inserted prompt is formulated as:
\begin{equation}
    P_l =
    [P_l^{ada}; p^V]
    =
    \left[
    \sum_{k=1}^{K} \alpha_{l,k} B_{l,k};
    W_I g^V
    \right].
\end{equation}
Here, $W_I$ is a learnable projection matrix, $P_l^{ada}\in\mathbb{R}^{B\times M\times d}$ denotes the routing-guided adaptive prompt, and $p^V\in\mathbb{R}^{B\times 1\times d}$ denotes the visual instruction token. 
Therefore, the final prompt satisfies $P_l \in \mathbb{R}^{B \times (M+1) \times d}$.

The adaptive prompt $P_l^{ada}$ provides sample-specific and layer-specific prompt modulation, while $p^V$ explicitly injects compact visual affective evidence into the textual encoding process. 
Together, they enable the model to handle diverse text-visual affective relations through controlled visual guidance.

\subsection{Training Objective and Parameter-efficient Tuning}
\label{subsec:objective}

\paragraph{Training objective}

For sentiment intensity prediction, we optimize the model with a standard regression objective. 
Given the predicted sentiment score $\hat{y}$ and the ground-truth score $y$, the task loss is defined as:
\begin{equation}
    \mathcal{L}_{task} = |\hat{y} - y|.
\end{equation}
For classification settings, $\mathcal{L}_{task}$ can be replaced by the cross-entropy loss.

Since the adaptive prompt is composed from a trainable prompt basis bank, the router may collapse to a small subset of prompt bases, leaving other bases insufficiently activated and trained. 
To alleviate this issue, we introduce a prompt activation balance loss. 
Let $\alpha_{l,b,k}$ denote the routing weight of the $k$-th prompt basis for the $b$-th sample at the $l$-th prompted layer. 
We first compute the average activation of each prompt basis within a mini-batch:
\begin{equation}
    m_{l,k} = \frac{1}{B} \sum_{b=1}^{B} \alpha_{l,b,k},
\end{equation}
where $B$ is the batch size. 
The balance loss for the $l$-th layer is formulated as:
\begin{equation}
    \mathcal{L}_{bal}^{l} = K \sum_{k=1}^{K} m_{l,k}^{2} - 1.
\end{equation}
The balance loss reaches its minimum when different prompt bases are uniformly utilized at the batch level. 
The final balance loss is averaged over all prompted layers:
\begin{equation}
    \mathcal{L}_{bal} =
    \frac{1}{|\mathcal{S}|}
    \sum_{l \in \mathcal{S}} \mathcal{L}_{bal}^{l},
\end{equation}
where $\mathcal{S}$ denotes the set of prompted layers. 
The overall training objective is:
\begin{equation}
    \mathcal{L} =
    \mathcal{L}_{task}
    + \lambda_{bal}\mathcal{L}_{bal},
\end{equation}
where $\lambda_{bal}$ controls the strength of the prompt balance regularization.

\paragraph{Parameter-efficient tuning}
To preserve pretrained knowledge and reduce training cost, we freeze the visual encoder and the BERT backbone during training. 
The visual encoder is used only for guidance feature extraction, while BERT serves as the text semantic backbone. 
Only the prompt-related and task-specific modules are trainable, including the prompt basis bank, co-guided routing projections, visual instruction projection, and sentiment prediction head:
\begin{equation}
    \Theta_{\mathrm{train}} =
    \{\Theta_{\mathrm{prompt}},
    \Theta_{\mathrm{router}},
    \Theta_{\mathrm{inst}},
    \Theta_{\mathrm{head}}\}.
\end{equation}
The parameters of the visual encoder and BERT backbone remain frozen:
\begin{equation}
    \Theta_V, \Theta_B \notin \Theta_{\mathrm{train}}.
\end{equation}

In this way, VG-TPT injects fine-grained visual affective cues into text encoding without full-model fine-tuning.

\FloatBarrier

\section{Experiments}
\label{sec:experiments}

\subsection{Datasets and Evaluation Metrics}
\label{subsec:datasets}

We evaluate VG-TPT on two widely used multimodal sentiment analysis benchmarks, CMU-MOSI and CMU-MOSEI~\cite{zadeh2016mosi,zadeh2018mosei}.
Both datasets consist of opinion utterances collected from online videos, where each utterance is annotated with a sentiment intensity score ranging from $-3$ to $+3$.
In this work, we focus on the vision-to-text setting, where the textual modality serves as the semantic anchor and the visual modality provides fine-grained affective guidance.

For CMU-MOSEI, we use 17,016 utterances for training, 1,813 for validation, and 4,098 for testing.
For CMU-MOSI, we use 1,283 utterances for training, 214 for validation, and 686 for testing.
Unlike many previous methods that directly use pre-aligned low-level multimodal features, our visual branch extracts guidance information from raw video frames sampled from the original videos.
Specifically, we sample the middle frame of each utterance-level video segment to obtain the visual guidance representation.

Following common practice, we formulate sentiment prediction as a regression task and report Mean Absolute Error (MAE), seven-class accuracy (Acc-7), binary accuracy (Acc-2), and F1 score.
MAE evaluates the error of continuous sentiment prediction, while Acc-7 is computed by mapping continuous predictions to seven discrete sentiment levels.
Acc-2 and F1 are used to evaluate binary sentiment polarity prediction.

\subsection{Main Results}
\label{subsec:main_results}

We compare VG-TPT with three groups of baselines.
First, text-only baselines are included to evaluate the strength of the semantic anchor.
Second, a visual-only Transformer is reported to show the limited predictive ability of visual information alone.
Third, mainstream full-modality methods using acoustic, visual, and textual modalities are included as reference results~\cite{williams2018eflstm,tsai2019mult,han2021mmim}.
It should be noted that these full-modality methods use additional acoustic information and are therefore not strictly comparable to our V+T setting.

\begin{table}[!htbp]
\centering
\caption{
Main results on CMU-MOSEI. 
A, V, and T denote acoustic, visual, and textual modalities, respectively. 
$\uparrow$/$\downarrow$ indicates that higher/lower values are better.
Boldface denotes the best performance, and underlining denotes the second-best performance.
}
\label{tab:mosei_results}
\setlength{\tabcolsep}{3.5pt}
\begin{tabular}{lccccc}
\toprule
Method & Mod. & Acc-7 $\uparrow$ & Acc-2 $\uparrow$ & F1 $\uparrow$ & MAE $\downarrow$ \\
\midrule
BERT~\cite{liu2025dstem} & T & 49.22 & 82.12 & 82.81 & \textbf{0.557} \\
XLNet~\cite{liu2025dstem} & T & 51.16 & \underline{83.32} & \underline{83.09} & 0.569 \\
V-Transformer~\cite{tsai2019mult} & V & 43.5 & 66.4 & 69.3 & 0.759 \\
\midrule
EF-LSTM~\cite{williams2018eflstm} & A+V+T & 46.3 & 76.1 & 75.9 & 0.594 \\
MulT~\cite{tsai2019mult} & A+V+T & 50.7 & 81.6 & 81.6 & 0.591 \\
MMIM~\cite{han2021mmim} & A+V+T & \textbf{52.6} & 81.5 & 81.3 & 0.578 \\
\midrule
\textbf{VG-TPT} & V+T & \underline{52.2} & \textbf{84.6} & \textbf{84.5} & \underline{0.565} \\
\bottomrule
\end{tabular}
\end{table}

\begin{table}[!htbp]
\centering
\caption{Main results on CMU-MOSI. Notation is consistent with Table~\ref{tab:mosei_results}.}
\label{tab:mosi_results}
\setlength{\tabcolsep}{3.5pt}
\begin{tabular}{lccccc}
\toprule
Method & Mod. & Acc-7 $\uparrow$ & Acc-2 $\uparrow$ & F1 $\uparrow$ & MAE $\downarrow$ \\
\midrule
BERT~\cite{liu2025dstem} & T & 42.17 & 83.60 & 84.80 & \underline{0.731} \\
XLNet~\cite{liu2025dstem} & T & 44.89 & \textbf{84.52} & \textbf{86.74} & \textbf{0.723} \\
V-Transformer~\cite{tsai2019mult} & V & -- & -- & -- & -- \\
\midrule
EF-LSTM~\cite{williams2018eflstm} & A+V+T & 31.0 & 73.6 & 74.5 & 1.420 \\
MulT~\cite{tsai2019mult} & A+V+T & 39.1 & 81.1 & 81.0 & 0.889 \\
MMIM~\cite{han2021mmim} & A+V+T & \underline{45.9} & 83.4 & 83.4 & 0.777 \\
\midrule
\textbf{VG-TPT} & V+T & \textbf{46.1} & \underline{83.7} & \underline{83.6} & 0.753 \\
\bottomrule
\end{tabular}
\end{table}

As shown in Table~\ref{tab:mosei_results}, VG-TPT achieves strong performance on CMU-MOSEI under the V+T setting.
Compared with text-only baselines, VG-TPT obtains higher Acc-7, Acc-2, and F1, indicating that visual affective cues provide complementary information for text-centered sentiment understanding.
Although full-modality methods use an additional acoustic modality, VG-TPT still achieves competitive or superior performance on several metrics.

Table~\ref{tab:mosi_results} reports the results on CMU-MOSI.
VG-TPT achieves the best Acc-7 among the compared methods and obtains competitive results on Acc-2, F1, and MAE.
These results indicate that VG-TPT performs competitively on the smaller CMU-MOSI dataset.

\subsection{Ablation Study}
\label{subsec:ablation}

To verify the contribution of each component, we conduct ablation studies on CMU-MOSEI.
We consider four variants of VG-TPT.
\textit{w/o Adaptive Prompt} replaces the dynamically generated adaptive prompt with fixed layer-wise prompts.
\textit{w/o Layer-wise Prompt Updating} generates the co-guided prompt only at the first prompted layer and shares it across subsequent layers.
\textit{w/o Text State} removes the current text state from the router and uses only visual guidance to generate routing scores.
\textit{w/o Prompt Selection} removes selective prompt activation and uniformly averages all prompt bases.

\begin{table}[!htbp]
\centering
\caption{Ablation study on CMU-MOSEI. Boldface denotes the best result, and underlining denotes the second-best result.}
\label{tab:ablation}
\setlength{\tabcolsep}{4pt}
\begin{tabular}{lcccc}
\toprule
Method & Acc-7 $\uparrow$ & Acc-2 $\uparrow$ & F1 $\uparrow$ & MAE $\downarrow$ \\
\midrule
w/o Adaptive Prompt & 50.6 & 83.4 & 83.2 & 0.586 \\
w/o Layer-wise Updating & 51.1 & 83.8 & 83.7 & 0.578 \\
w/o Text State & 50.9 & 83.6 & 83.4 & 0.582 \\
w/o Prompt Selection & 51.4 & 84.0 & 83.8 & 0.574 \\
\midrule
\textbf{Full Model} & \textbf{52.2} & \textbf{84.6} & \textbf{84.5} & \textbf{0.565} \\
\bottomrule
\end{tabular}
\end{table}

As shown in Table~\ref{tab:ablation}, removing the adaptive prompt leads to a clear performance drop, indicating that fixed prompts are insufficient to model diverse text-visual affective relations.
Sharing the generated prompt across layers also degrades the performance, which verifies the necessity of layer-wise prompt updating.
When the current text state is removed from the router, the model cannot select prompts according to evolving textual semantics, resulting in inferior performance.
Moreover, replacing selective prompt activation with uniform prompt fusion weakens the model, demonstrating the importance of routing-guided prompt composition.

\subsection{Parameter Efficiency}
\label{subsec:param_eff}

To further evaluate the parameter efficiency of VG-TPT, we compare the number of trainable parameters with full BERT fine-tuning and a representative full-modality baseline.
Although VG-TPT introduces vision-guided prompt modules, most pretrained parameters remain frozen.
Only the prompt basis bank, routing projections, visual instruction projection, and prediction head are updated during training.

\begin{table}[!htbp]
\centering
\caption{
Parameter efficiency comparison on CMU-MOSEI.
Params denotes the number of trainable parameters in millions (M), and Trainable Ratio denotes the percentage of trainable parameters relative to the total parameters of each model.
Bold indicates VG-TPT results.
}
\label{tab:param_eff}
\setlength{\tabcolsep}{5pt}
\begin{tabular}{lcccc}
\toprule
Method & Params (M) & Ratio & Acc-7 $\uparrow$ & Acc-2 $\uparrow$ \\
\midrule
BERT full fine-tuning~\cite{liu2025dstem} & 109 & 100\% & 49.22 & 82.12 \\
MulT~\cite{tsai2019mult} & -- & 100\% & 50.7 & 81.6 \\
\textbf{VG-TPT} & \textbf{2.4} & \textbf{1.2\%} & \textbf{52.2} & \textbf{84.6} \\
\bottomrule
\end{tabular}
\end{table}

As shown in Table~\ref{tab:param_eff}, full BERT fine-tuning updates all BERT parameters, whereas VG-TPT updates only 2.4M parameters, accounting for 1.2\% of its total model parameters.
Compared with full BERT fine-tuning, the number of trainable parameters is substantially reduced.
Meanwhile, VG-TPT achieves better Acc-7 and Acc-2 than BERT full fine-tuning, showing that it can effectively incorporate visual affective cues with only a small number of trainable parameters.

\FloatBarrier

\section{Conclusion}
\label{sec:conclusion}

In this paper, we propose VG-TPT, a parameter-efficient vision-guided text prompt tuning framework for multimodal sentiment analysis. 
VG-TPT injects visual facial cues into the BERT textual encoder through layer-wise adaptive prompts, enabling visual information to refine and disambiguate text-centered sentiment representations. 
A co-guided router jointly uses the current text state and visual guidance feature to selectively compose adaptive prompts from a trainable prompt bank. 
Experiments on CMU-MOSI and CMU-MOSEI show that VG-TPT achieves competitive performance with only 2.4M trainable parameters. 
Future work will extend VG-TPT to audio-visual-text chain-guided sentiment modeling and more fine-grained temporal visual reasoning.

\bibliographystyle{IEEEtran}
\balance
\bibliography{mmsp_refs}

\end{document}